\documentstyle[12pt,aasms4]{article}

\begin{document}

\title{Average Emissivity Curve of BATSE Gamma-Ray Bursts with Different
Intensities}
\author{Igor G.~Mitrofanov$^1$, Maxim L.~Litvak$^1$
\and %
Michael S.~Briggs$^2$, William S.~Paciesas$^2$, Geoffrey N.~Pendleton$^2$,
Robert D.~Preece$^2$ \and Charles A.~Meegan$^3$}
\affil{$^1$Space Research Institute, Profsojuznaya str. 84/32, 117810
Moscow, Russia}
\affil{$^2$Department of Physics, University of Alabama in Huntsville
Huntsville, AL 35899}
\affil{$^3$NASA/Marshall Space Flight Center, Huntsville, AL 35812}

\begin{abstract}
Six intensity groups with $\sim 150$ BATSE gamma-ray bursts each are compared
using average emissivity curves. Time-stretch factors for each of the dimmer
groups are estimated with respect to the brightest group, which serves as the
reference, taking
into account the systematics of counts-produced noise effects and choice
statistics. 
A stretching/intensity anti-correlation is found
with good statistical significance during
the average back slopes of bursts. A stretch
factor $\sim 2$ is found between the 150 dimmest bursts, with peak flux $<0.45$
ph cm$^{-2}$ s$^{-1}$, and the 147 brightest bursts, with peak flux $>4.1$
ph cm$^{-2}$ s$^{-1}$. On the other hand, while a trend of increasing stretching factor
may  exist for rise fronts for burst with decreasing peak flux from $>4.1$
ph cm$^{-2}$ s$^{-1}$ down to $0.7$ ph cm$^{-2}$ s$^{-1}$, the magnitude of the stretching
factor is less than  $\sim$ 1.4 and is therefore inconsistent with stretching factor
of back slope.

\end{abstract}

\section{Introduction}

Gamma-ray bursts are known to have very different intensities. The peak flux
of bursts varies more than 2 orders of magnitude from the BATSE
trigger threshold of
about 0.3 ph cm$^{-2}$ s$^{-1}$ up to the highest measured values of about 50
ph cm$^{-2}$ s$^{-1}$ (\cite{Fish}). The rich statistics of the BATSE 4B
Catalog
(\cite{Pacia}) enables one to divide all detected bursts into several
intensity groups with reasonably large numbers of events ($\sim 150$) in each
of them. Although individual bursts have very different time profiles, a
characteristic average temporal signature can be produced for each group.
Using these average signatures, a comparison can be made
between the duration of different brightness groups of gamma-ray bursts. The
main goal of this comparison is to test the brightness-dependent stretching
of gamma-ray bursts.

It is well-known that individual pulses from radio pulsars have quite variable
emission time
profiles, so that the characteristic periodic pulsar signal is hardly
recognizable by examining a short interval of real-time data. The randomized
time profile is transformed into the stable signature of a pulsar light
curve by using the epoch folding technique, averaging the data over many
periods. A similar signature has been proposed for cosmic
gamma-ray bursts: averaging the time profiles of individual events by
the normalized peak-alignment technique, where each time profile is normalized
by the peak number of counts $C_{\rm max}$, aligned at peak time bins
$t_{\rm max}$
and then averaged for all bins along the time scale (Mitrofanov et al. 1996,
hereafter \cite{MitrI}). This
technique produces an {\it Average Curve of Emissivity}, or ACE, and the
corresponding signature represents the averaging of observed time histories
of bursts.

ACE curves have already been studied for different energy ranges and for
different intensity groups of bursts, and they have been found to be rather
convenient signatures to describe the basic properties of the slow temporal
variation of bursts (\cite{MitrII}). This signature averages out the fast
variations of bursts in the sample and describes the general envelope of
rising emission before the main peaks of bursts and the subsequent decaying
tail. An analytic approximation to the ACE has been found, which results in
a very acceptable fit to the observations. The ACE {\it equivalent width}
parameter describes the mean time scale of the slow variation of GRBs.

The comparison of the generic time-dependent properties of gamma-ray bursts with
different brightnesses has recently become an issue of common interest.
Indeed, if the difference in intensities represents the difference in
distances to the emitters, then the difference between the average time
signatures of different brightness groups should manifest effects of
cosmological
time-dilation, where dimmer bursts are observed to be broader than brighter
events
because they were emitted at larger cosmological distances. The
time-dilation test based on the comparison of the averaged temporal structures
for dim and for bright sets has already been done by two groups, which have
drawn two opposite conclusions: in the first case a large time-dilation effect
$\sim 2$ was seen (\cite{Norra,Norrb}), in the other case none was observed
(\cite{MitrIII} \& \cite{MitrI}).

The effect of cosmological time-dilation should also be accompanied by
the effect of photon energy red-shift, which could influence the
time-stretching
value (\cite{Fenima,MitrII}). On the other hand, sources of bright and dim
gamma-ray bursts could be non-standard candles with intrinsic
luminosity-based correlations (\cite{Brain}). In this case, the observed
time-stretching of dim bursts could result from the fact that shorter
outbursts correlate with smaller intrinsic luminosity of their sources.
Also, one should take into account possible distance-related evolution of
bursts emitters: in the co-moving reference frames, more distant sources could
have shorter time profiles than nearer sources. These intrinsic effects
could interfere with the cosmological effects.

Therefore, in order to make conclusive statements about generic time-dependent
properties of gamma-ray bursts, one should distinguish between the physical
effect
of time-dilation, which results from the expansion of the Universe, and the
phenomenological effect of time-stretching, which is found by the comparison
of bursts with different intensities. We think that the first logical step
should be the investigation of the stretching/intensity phenomenon by
comparing time profiles for different intensity groups of bursts.
If a significant time-stretching effect is found between different intensity
groups, then the next logical step can be made. This step should be the
physical interpretation of the observed stretching, according to predictions of
a cosmological model, taking into account the geometrical effects of
time-dilation and red-shift, the physical effects of time-energy dependence
of emission, the effects of a broad luminosity distribution of emitters, and
finally, the effects of distance-related evolution.
In this paper, we restrict ourself to purely phenomenological studies of
the stretching/intensity correlation of bursts, using the average emissivity
curves for different brightness groups of bursts as the tool for measuring
the stretching.

A total set of 887 bursts with $t_{90}>2$ s from the BATSE 4B Catalog
(\cite{Meega,Pacia}) have
data available on the 1024 ms timescale. We have divided these into 6 intensity
groups (Table 1), using the peak flux parameter $F_{\rm max}^{(1024)}$ as the
selection criterion. For each of these groups, an ACE profile is produced
using counts observed in the broad energy range 50-300 keV. The technique used
to produce ACE profiles is described in \cite{MitrI}.
Below, the stretching/intensity effect is investigated and stretching
factors are evaluated between each ACE$_i$ for the dimmer groups ($i=2$ --
6) and
the reference group (ACE$_1$). We have done this analysis for the total profile
as well as for the rise fronts and the back slopes separately.

\section{Analytical approximation of ACE profiles}

We have previously determined that there is a simple analytical form that
fits the ACE profiles of all burst intensity groups in several
different spectral ranges quite well (\cite{MitrII}).
For each selected group ($i$) of bursts, the analytical approximation to
the ACE is
\begin{equation}
f^{(i)}[t]=\left(\frac{\Delta t^{(i)}}{\Delta t^{(i)}+|t - t_{\rm
max}|}\right)^{a^{(i)}_{\rm RF},
a^{(i)}_{\rm BS}} {\rm ,}
\end{equation}
\noindent with different power indices ($a_{\rm RF}^{(i)}$, $a_{\rm
BS}^{(i)}$)
at the rise front (RF, with $t<t_{\rm max}$) and at the back slope (BS, with
$t>t_{\rm max}$), respectively.

Another approximation for ACE-like profiles with exponential wings has been
suggested by \cite{Stern}. A direct comparison between these two approximations
has shown that a power law (eq.~1) fits the 1024 ms time scale ACE profiles
over the range of 20 time bins before and after the peak much better than the
exponential model. For example, with intensity group 1 (Table 1) we found a
reduced $\chi ^2=1.51$ for an exponential-type
law, while Eq.~1 provided a better fit, with reduced $\chi ^2=0.88$.
For this reason, we use Eq.~1 to model ACE profiles in this paper.

\section{Two procedures to estimate stretch factors between ACE curves of
different intensity groups}

When cosmological models were first suggested for GRBs, the observational
tests for cosmological signatures were expected be quite obvious; for the
case of ACE stretching coefficients, a factor of at least 2 should be
observed between the brightest and dimmest burst intensity groups.
Therefore, a simple procedure
was devised to look for evidence for this stretching (see \cite{MitrI}).
However, a stretching factor of the expected magnitude was
not found; indeed, recently-developed cosmological models for bursts predict
much smaller average stretching, despite the fact
that some emitters should exist at large cosmological red-shifts
(see e.g. \cite{Brain}). In order to make further progress, a more sensitive
comparison method between different ACE profiles has been developed,
allowing the measurement of weak-signature stretching factors 
(less than 1.5). To be successful, this method must take into account several
statistical biases associated with the limited number of bursts in different
intensity group samples and with the poor counts statistics of weak events.

We have developed two such procedures to
make a robust comparison of ACE profiles between different brightness
groups of bursts.
Procedure (a), which is relatively time-consuming, requires the following
steps to
compare  intensity groups $i=2$ -- 6 with the reference group
$i=1$:

{\bf (a1)} An arbitrary stretching coefficient $Y$ is selected, and each
burst of the reference group 1 is stretched by the factor $Y$.

{\bf (a2)} An artificially-dimmed reference group, which has the same
signal-to-noise ratio as group $i$, is created from bursts of
the stretched reference group (see Section 5).

{\bf (a3)} ACE profiles are produced for the 
stretched and dimmed version of the
reference group and
for the original group $i$. The value of $\chi ^2$ is used to measure
the difference between them. To find the best fit stretching coefficient
$Y_i$, the value of the
stretching coefficient $Y$ is changed, and the cycle (a1)--(a3) is
performed again to minimize $\chi^2$.

The time-efficient procedure (b) is more simple and straight-forward
than the first one. It consists of the following steps:

{\bf (b1)} An artificially-dimmed reference group that has equal
signal-to-noise ratio to the the  group $i$ is created from the
reference group (see Section 5).

{\bf (b2)} An ACE profile is produced for the artificially-dimmed reference
group, and the parameters of the best-fit analytical model (Eq.~1) are
computed.

{\bf (b3)} The stretching parameter Y for the time constant $\Delta t^{(1)}$
is introduced
into the analytical model (Eq.~1) for the artificially dimmed reference
group 1:

\begin{equation}
f_{\dim }^{(i)}[t]=\left(\frac{Y_{i}\cdot \Delta t^{(1)}}{Y_{i}\cdot
\Delta t^{(1)}+|t-t_{\rm max}|}\right)^{a_{\rm RF}^{(1)},a_{\rm BS}^{(1)}} .
\end{equation}

\noindent Fixing all other parameters, the best fit value for $Y$, as a
free parameter, is found for the ACE of each  group $i=2-6$. This
value is taken to be the stretch coefficient $Y_i$ between the  group
$i$ and the reference group.

A direct comparison of these two procedures has been performed for the
back slopes (time bins after the peak in the ACE) of bursts using the 3B
Catalog data base (\cite{Litv1}). Excellent agreement between the best-fit
stretch factors ($Y$) estimated from both procedures was found.
The difference between them is much less than 1$\sigma$ for choice
statistics (see Section 4). Therefore, the computationally time-efficient
procedure, (b), is used below in studies of ACE stretching.

\section{Choice statistics and the related errors of the
stretching coefficients}

The proposed method suggests that we may use the ACE as the
signature representing the slow
variability  of a selected sample of gamma-ray bursts. Individual
bursts have very different time profiles, and one must use a
very large number of individual events to build up a representative ACE
that will be the  same for different  samples of bursts
with similar brightnesses.

The errors associated with ACE curves may be estimated from the
sample variance
for the selected groups of bursts. ACE profiles for two independent
groups are statistically indistinguishable, provided the difference
between them is within these errors. The direct comparison of ACE profiles
for different groups has shown that they are much more different than
might expected from the variance within each sample only. This means that
these groups are not representative samples, and a random selection of
bursts within these groups does not ensure well-weighted contributions
from all kinds of possible profiles.

To study the {\it random choice statistics} of bursts, a Monte
Carlo test has been performed using the total set of 603 bursts from the 3B
Catalog (\cite{Meegb}) that have catalog values of peak flux
and durations $T_{90}>2$ s (\cite{Litv2}). Different numbers of bursts,
$N=100$, 150, 200, 250, 300, were assumed. We randomly selected
$10^4$ times, two testing groups of $N$ events each from the
total set of bursts, produced ACEs for these groups, and then determined the
best-fit stretch factor $Y_{\rm choice}$ between them. The distributions
of $Y_{\rm choice}$ for two different values for $N$ are presented in Fig.~1.

The spread of stretching coefficients $Y_{\rm choice}$ due to random choice
statistics is found to be much broader than would be expected from the
sample variance for each individual group. Of course, the distribution of
$Y_{\rm choice}$ becomes narrower for larger $N$, but even for the largest
value, $N=300$, it is still quite broad.

Therefore, we conclude that for any two intensity groups, with $N$ bursts
each, the estimated stretching coefficient must be compared with standard
deviations from the random choice distribution, evaluated for $N$ events. For
$N=150$, the level of 1$\sigma$ significance corresponds to the stretching
coefficient $\sim $1.10. Therefore, for $Y\sim 1$, one should use the errors $%
\delta Y\sim 0.10$ for the 1$\sigma$ errors of the stretching coefficients.
In the arbitrary case where $N \geq 100$, one can use following for
estimations of errors:
\begin{equation}
\frac{\delta Y}Y=0.10\cdot \sqrt{ \frac{150}N } .
\end{equation}
\noindent Below, these errors were used to estimate the significance of
stretching between selected brightness groups (Tables 2 and 3).

\section{Count-noise produced effects on the ACE}

The procedure to create each ACE includes the selection of the highest peak
of each burst, with counting rate $C_{\rm max}$, and the normalization of
the time profile by
the $C_{\rm max}$ value. Therefore, the ACE is sensitive to a bias
where positive fluctuations of counts ($C+\delta C$)
dominate the selection of $C_{\rm max}$. When the normalization is
performed, a profile is lowered $(C_{\rm max}+\delta C)/C_{\rm max}$ times.
The effect should be larger for dimmer bursts, where the ratio of
$(C_{\rm max}+\delta C)/C_{\rm max}$ is larger.

To evaluate the effect, the reference group 1 (Table 1) has been transformed
by the procedure of Monte Carlo noisification into 5 artificial reference
groups, which have peak flux distributions similar to the corresponding
distributions for the 5 observed dim intensity groups, $i=$2--6 (Table 1). The
Monte Carlo transformation procedure includes the following steps:

a) For each original burst of the reference group (1), some counterpart event
is randomly selected inside the testing group $i$.

b) The ratio of peak fluxes $f=F_{\max }^{(1)}/F_{\max }^{(i)}$ is estimated
between the peak fluxes of the original burst and its weaker counterpart.

c) The time profile of the original bright burst is divided by the factor $f$,
and counts for an artificially-dimmed version of the
original burst are simulated using Poisson statistics:
\begin{equation}
D_j^{}=\frac{C_{{\rm S},j}}f+C_{{\rm B},j},
\end{equation}
where $C_{{\rm S},j}$, $C_{{\rm B},j}$ are the signal and background counts
accumulated during the $j$th time bin for a burst from the reference group.

Using equation (2), stretching coefficients $Y_{\rm noise}$ have been
estimated between the ACE$_1$ for the original reference group and
ACE$_i^{\rm art}$ for the artificial reference groups ($i=2$--6).

One can see the largest effect is that the ACE$_6^{\rm art}$ for the
dimmest group
($i=6$) must be broadened by a factor of 1.22 to equal the width of the
ACE$_1$
for the original reference group. 
Therefore,
if the observed stretching factor between ACEs for the actual observed dim
group and the reference group is equal to 1, it should be interpreted
as evidence for real stretching on the order of $Y_{\rm noise}$ between the
dim and
reference groups, because the ACE for the dim group is known to be narrowed
systematically due to larger relative Poisson noise. This effect could be
ignored
when testing for much larger stretching coefficients above $2.0$
(\cite{MitrI}),
but it should certainly be taken into account when the expected stretching
might be as
small as the noise-produced effect (see below). Both procedures (a) and (b)
(Section 3) take this effect into account.

\section{Stretching factors between the ACE for different intensity groups}

\subsection{Stretching with respect to the reference group}

Using procedure (b) to compare between ACEs, stretching
coefficients were estimated for each intensity group (Figure 2 and
Table 2). The group of brightest events (1) is used as a ${\it %
reference}$ and the corresponding stretching coefficient for this group was
defined to be  1. 
For each group, the stretching factors $Y_{\rm RF}^{(1)}$,
$Y_{\rm BS}^{(1)}$ and $Y_{\rm TOT}^{(1)}$ were estimated with respect to
group 1  for the RF
and BS wings and for the total ACE profiles, respectively (Figure 2 and Table 2)
.

The rise front portions of bursts do not manifest any
significant increase of stretching with decreasing intensity of bursts.
The stretch factor for the rise front of group 5 with respect to the reference
group is the only one that is above the 3$\sigma $ level (in this
case, 3$\sigma = 1.34$). On the other hand, at the back
slopes the stretching factors $Y_{\rm BS}^{(1)}$ are larger than $3\sigma$
already between groups 2 and 1, and increase up to $\sim $2 with decreasing
brightness (Table 2). The noise-produced systematic narrowing of the ACE is
taken
into account for these values, because the test groups ($i=2$ -- 6) are
compared with the correspondingly noisified reference groups. In the case of
the comparison with the original reference group (1), the stretching would
be less. The lack of stretching  during the rising portion of the ACE of the
dimmest group (6) may be an instrumental effect,
resulting from the deficit of slow-rising events near the trigger threshold
(e.~g. \cite{Higdon}). We know from the non-triggered burst sample of
\cite{Kommers}
that the threshold effect may change the estimate for the rise-front
stretching
factor only for the dimmest group 6 in our sample.

When the total ACE profiles are compared, the stretching factors have
intermediate values between the corresponding factors for rise fronts
and back slopes. One can see that the back slope stretching factors
for dimmer groups ($i=3$--5) are $\sim 2 \sigma$ larger than
those associated with the rise front (Table 2). That is why one should
study the stretching phenomena of bursts 
separately for the rise front and back slopes.

\subsection{Stretching with respect to the second brightest group}

With decreasing burst intensities, the largest increase of
stretching between successive intensity groups happens between the brightest
and the second brightest groups (Table 2 and Figure 1). Of course,
this could result from a random choice fluctuation: the number of
bursts in each group is still not large enough to exclude this possibility.
Larger
burst samples from subsequent BATSE catalogs should be used to check the
jump of
stretching for the two intensity groups bounded by the flux of
$\sim 4.1$ ph cm$^{-2}$ s$^{-1}$. The jump may
completely disappear for larger statistics; or, more interestingly, it could
be confirmed as a real phenomenological effect.

However, one has to be sure that this jump-like effect between the test
group ($i=2$) and the reference ($i=1$) does not result from some systematic
effect in the comparison procedure. Indeed, the reference group has been
transformed by adding noise in order to be to be compared with dimmer
groups, and some
unknown systematics in this procedure could result in a jump of stretching
between the reference group and dimmer groups.

To check this possibility, the same procedure used for the
estimations of stretching factors has been reproduced; however, this time
assuming the second bright group (2) to be the reference (Table
3). The stretching factors $Y_{\rm RF}^{(2)}$ and $Y_{\rm BS}^{(2)}$ for
group 2
have been defined as 1.0 in this case. To facilitate comparison between
stretching coefficients based on group 1 and group 2, the
re-normalized factors $1.19\cdot Y_{\rm RF}^{(2)}$ and $1.35\cdot Y_{\rm
BS}^{(2)}$
are also presented in Table 3. There is very good agreement between
stretching factors $Y^{(1)}$ (Table 2) and $Y^{(2)}$ (Table 3), based on the
reference groups 1 and 2, respectively.

Therefore, one can exclude the possibility that some systematic effect takes
place when the reference group is compared with dimmer ones. One may
conclude that in the 4B catalog data set a large stretching factor of about
1.35
really exists between the back slopes of bursts of intensity groups 1 and 2,
separated by a flux $\sim 4.1$ ph cm$^{-2}$ s$^{-1}$.

\subsection{Consistency between stretching factors for intensity groups
of the 2B and 4B database}

A comparison between ACE profiles was done previously (\cite{MitrI})
using the database of 338 events from the BATSE 2B catalog (\cite{Meega}),
with durations $T_{90}>1$ s. These were divided at the peak flux value
$\sim 1$ ph cm$^{-2}$ s$^{-1}$ into two intensity groups with 143 bright
bursts and
179 dim bursts. The estimated equivalent time widths $t_{ETW}$ (see
\cite{MitrI}%
) of ACE rise fronts, back slopes and total profiles for these old samples
are presented in Table 4. The values labeled ``Old samples'' are taken from
\cite{MitrI} but the errors are estimated according to random choice statistics
as described above (Section 4). No significant stretching is seen between
the ACE
for these groups.

The 4B catalog allows us to do a similar analysis for new samples with as
many as 480 bright bursts and 464 dim events, and to check the consistency
between the estimations of equivalent time widths for the old and new samples.
One concludes from Table 4 that the ``old'' and ``new'' results are consistent
for intensity groups divided by the same peak flux value $\sim 1$ ph cm$^{-2}$
s$^{-1}$. There is no evidence for stretching for the ``old'' groups with
143 bright
and 179 dim events either during the rise front or the back slope. Using
the much better
statistics available with the ``new'' intensity groups, 
there
still is no significant stretching during the rise fronts, but
there is
some effect  when $t_{ETW}^{(\rm BS)}$ are compared.

The separation of all bursts into bright and dim groups by the peak flux
$\sim 1$ ph cm$^{-2}$ s$^{-1}$ was appropriate when the goal was to test
for an obvious signal of stretching by factors large as $\sim 2$. On the
other hand, the 2B catalog sample was too small for more accurate sampling.
Now, with much better statistics, more subtle stretching
effects can be seen for several intensity groups (Table 2), in particular
at the back slope of ACE profiles. In the half-to-half separation, a
stretching effect of $\sim 2 \sigma$ is also seen at the back slope
(Table 4).

\section{Conclusions}

Six burst intensity groups with $\sim 150$ events each have been
compared to determine a stretching effect between the ACE profiles for dim and
bright groups. To study the stretching, a separate comparison is preferable
for the average rise fronts and back slopes. The Pearson $\chi^2$ statistic
between the reference and the stretched dimmer groups
allows us to find the most probable stretching factors for the ACE of
different
intensity groups (Table 2), and the resulting reduced $\chi^2$ values are
significantly smaller when the rise fronts and back slopes are each compared
separately. There is a significant difference between the corresponding
stretching factors for the
rise fronts and back slopes in different intensity groups. During the rise
front a stretch factor of $1.39\pm 0.14$ is found between intensity
groups 5 and 1, which seems to be marginally significant. The other factors
determined for the rise fronts, taken together, also indicate some
stretching effect, but not significantly.

This study is based on BATSE 1~s resolution discriminator data.
Higher temporal resolution datatypes begin either at the trigger time or
2~s before.  Our choice of the continuously available 1~s data avoids
possible systematic effects in the rise front due to mixing datatypes 
of differing temporal resolutions.

There may be an instrumental triggering bias that selects against those dim
bursts that rise more slowly on the average. This has been corroborated
in the study of non-triggered bursts by \cite{Kommers}, who found that
only some of the dimmest bursts
failed to trigger due to slow rise times.  The intensities of these events
correspond to our 
dimmest intensity group 6.  
Since only group 6 is incomplete due to missing slow risers,
and since this incompleteness affects only the rise front results, we
base our conclusions on rise front stretching on groups 1 to 5.
Therefore, omitting
the dimmest group, there is the indication of a trend  for the
rise-front stretching factors (Figure 3).

At the back slopes the estimated stretch factors are
quite significant for all dimmer groups ($i=2$ -- 6) with respect to the
brightest one ($i=1$), and the largest factor between the ACE profiles of
dimmest 150 bursts and the brightest 147 bursts is about 2.1$\pm $ 0.2. The
non-stretching hypothesis may be significantly rejected for the back
slopes of bursts. Our results are in qualitative agreement with recent
estimations of time-stretching by \cite{Norra}, where a
different energy range was used for burst averaging and a different time
scale was applied for burst selection.

We conclude that models of gamma-ray bursts should 
explain two phenomenological results:

1) Back slopes for the 150 dimmest bursts, with peak fluxes $<0.45$ ph
cm$^{-2}$
s$^{-1}$, are on the average $\sim $2 times longer in duration than
the 150 brightest bursts, with peak fluxes $>4.1$ ph cm$^{-2}$ s$^{-1}$.

2) While a trend of increasing stretching factor, relative to group 1, may 
exist for rise fronts for groups 2 to 5, the magnitude of the stretching factor
is less than $\sim1.4$ and is therefore inconsistent with the stretching
factor of the back slope.

3)Finally, there is definitely no full-profile stretching between dim and bright
bursts as large as $\sim 2$ or more. 
Whatever stretching exists is weak and so the correct determination of
the stretching factor will require careful treatment of statistical and
systematic effects that are  comparable with the physical
effect
of stretching that we are looking for. Using the 4B data, we conclude that a
significant stretching by a factor of $\sim $2 between the different
brightness groups may only be resolved for the back slopes of the
average light curves  of GRB time profiles.

Our primary concern has been assessing the observational evidence for time
stretching in the average profiles of GRBs.
The observation of stretching in the back slopes but not in the rise fronts
of GRBs cannot be solely caused by cosmological effects -- at least one
of these phenomena seems to be intrinsic to GRBs, perhaps indicating a
slope correlation with absolute intensity or the existence of source
evolution.
Whatever the explanation, determining the distance scale of GRBs using
cosmological tests is proving to be more difficult than had
been hoped.

\section{Acknowledgments}

The work in USA was supported by NASA grant CRO-96-173. The work in Russia
was supported by RFBR grant 96-02-18825.

%

\clearpage

\figcaption{The $Y_{\rm choice}$ distribution for
N=150 events is shown by a thick line, while the distribution for N=300
events is
shown by a thin line.\label{Fig1}} 

\figcaption{The ACE for the reference group ({\it thick line}) is compared
with the
ACE for dim groups ({\it thin lines}).\label{Fig2}}

\figcaption{The best-fit stretching factors for the average back slopes
({\it solid})
and rise fronts ({\it dashed}) relative to the brightest
group ({\it solid arrow}) as a function of peak flux.
The value for the rise front for the dimmest group 6 may be altered by missing 
``slow riser'' triggers -- see text.\label{Fig3}}

\clearpage

\begin{deluxetable}{lccc}
\tablecaption{}
\label{table1}
\tablehead{
\colhead{Intensity} &
\colhead{Peak flux} &
\colhead{Number} &
\colhead{$Y_{\rm noise}$}\\
group & (ph cm$^{-2}$ s$^{-1}$) & of bursts}
\startdata
1 & $>$4.10    & 147 & 1.00 \nl
2 & 1.76-4.10 & 153 & 1.08 \nl
3 & 1.05-1.76 & 148 & 1.11 \nl
4 & 0.67-1.05 & 148 & 1.15 \nl
5 & 0.45-0.67 & 148 & 1.19 \nl
6 & $<$0.45    & 150 & 1.22 \nl
\enddata
\end{deluxetable}

\begin{deluxetable}{cccc}
\tablecaption{}
\label{table2}
\tablehead{
\colhead{Intensity group}&
&\colhead{Relative Stretching coefficients}&\\
&
\colhead{$Y_{\rm RF}^{(1)}$}&
\colhead{$Y_{\rm BS}^{(1)}$}&
\colhead{$Y_{\rm TOT}^{(1)}$}}
\startdata
1 & 1.00 & 1.00 & 1.00 \nl
2 & 1.19$\pm 0.12$ & 1.35$\pm 0.14$  & 1.28$\pm 0.13$  \nl
3 & 1.23$\pm 0.12$ & 1.54$\pm 0.15$  & 1.44$\pm 0.14$  \nl
4 & 1.22$\pm 0.12$ & 1.48$\pm 0.15$  & 1.39$\pm 0.14$  \nl
5 & 1.39$\pm 0.14$ & 1.78$\pm 0.18$  & 1.60$\pm 0.16$  \nl
6 & 1.07$\pm 0.11$ & 2.10$\pm 0.20$  & 1.50$\pm 0.15$  \nl
\enddata
\end{deluxetable}

\begin{deluxetable}{ccccc}
\tablecaption{}
\label{table3}
\tablehead{
\colhead{Intensity group}&
\colhead{$Y_{\rm RF}^{(2)}$} &
\colhead{$Y_{\rm RF}^{(2)}\cdot 1.19$} &
\colhead{$Y_{\rm BS}^{(2)}$} &
\colhead{$Y_{\rm BS}^{(2)}\cdot 1.35$}}

\startdata
2 &  1.00               &  1.19 & 1.00               & 1.35\nl
3 &  1.06$\pm 0.11$ & 1.26 & 1.16$\pm 0.12$ & 1.57 \nl
4 &  1.01$\pm 0.10$ & 1.20 & 1.11$\pm 0.11$ & 1.50 \nl
5 &  1.12$\pm 0.11$ & 1.33 & 1.30$\pm 0.13$ & 1.76 \nl
6 &  0.85$\pm 0.09$ & 1.01 & 1.52$\pm 0.15$ & 2.01 \nl
\enddata
\end{deluxetable}

\begin{deluxetable}{lcc}
\tablecaption{}
\label{table4}
\tablehead{
\colhead{ Parameters to compare}&
\colhead{Old samples} &
\colhead{New samples }}
\startdata
$t_{ETW}^{(\rm RF)}$ for bright groups &  2.47$\pm 0.27$  &  2.48$ \pm
0.15$ \nl
$t_{ETW}^{(\rm RF)}$ for dim groups   &  2.25$\pm 0.29$  &  2.35$ \pm 0.14$ \nl
\tableline
$t_{ETW}^{(\rm BS)}$ for bright groups &  4.16$\pm 0.46$  &  3.62$ \pm
0.21$ \nl
$t_{ETW}^{(\rm BS)}$ for dim groups   &   4.32$\pm 0.48$  &  4.27$ \pm
0.26$ \nl
\enddata
\end{deluxetable}

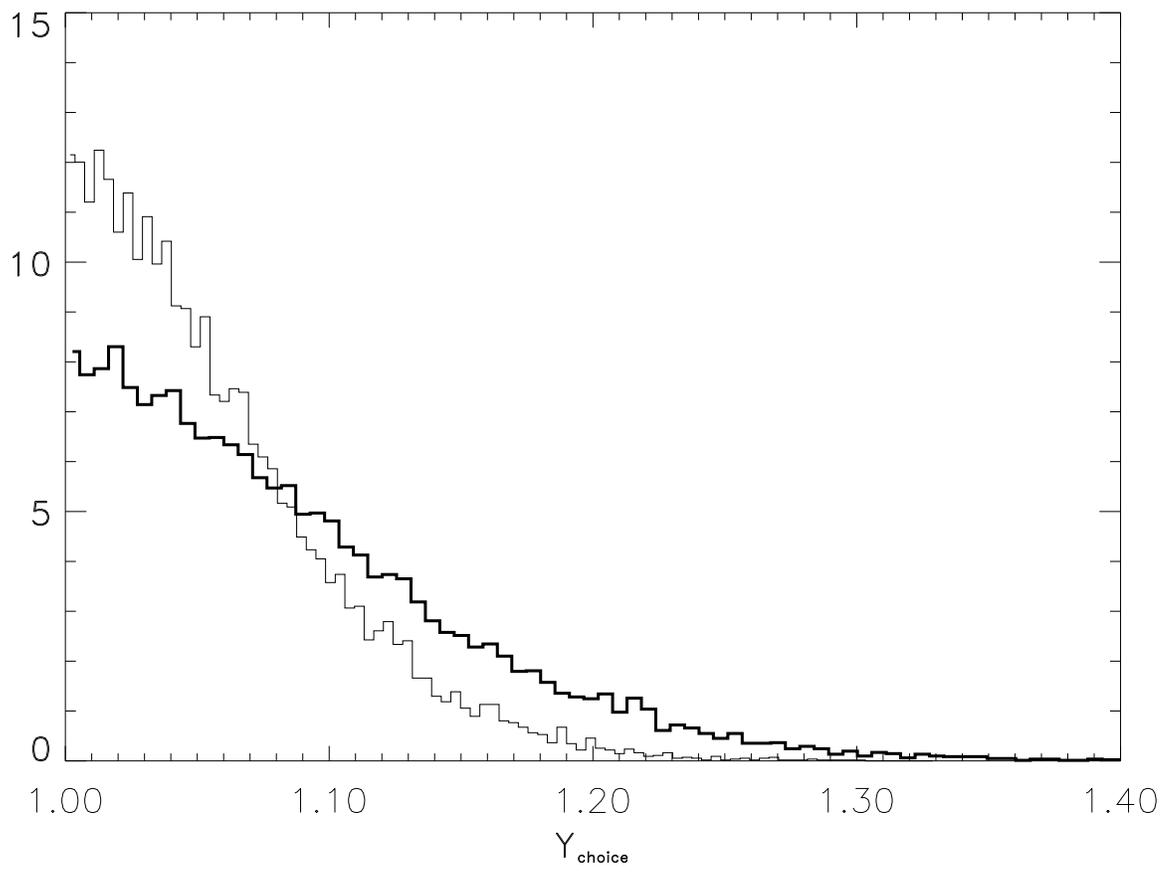
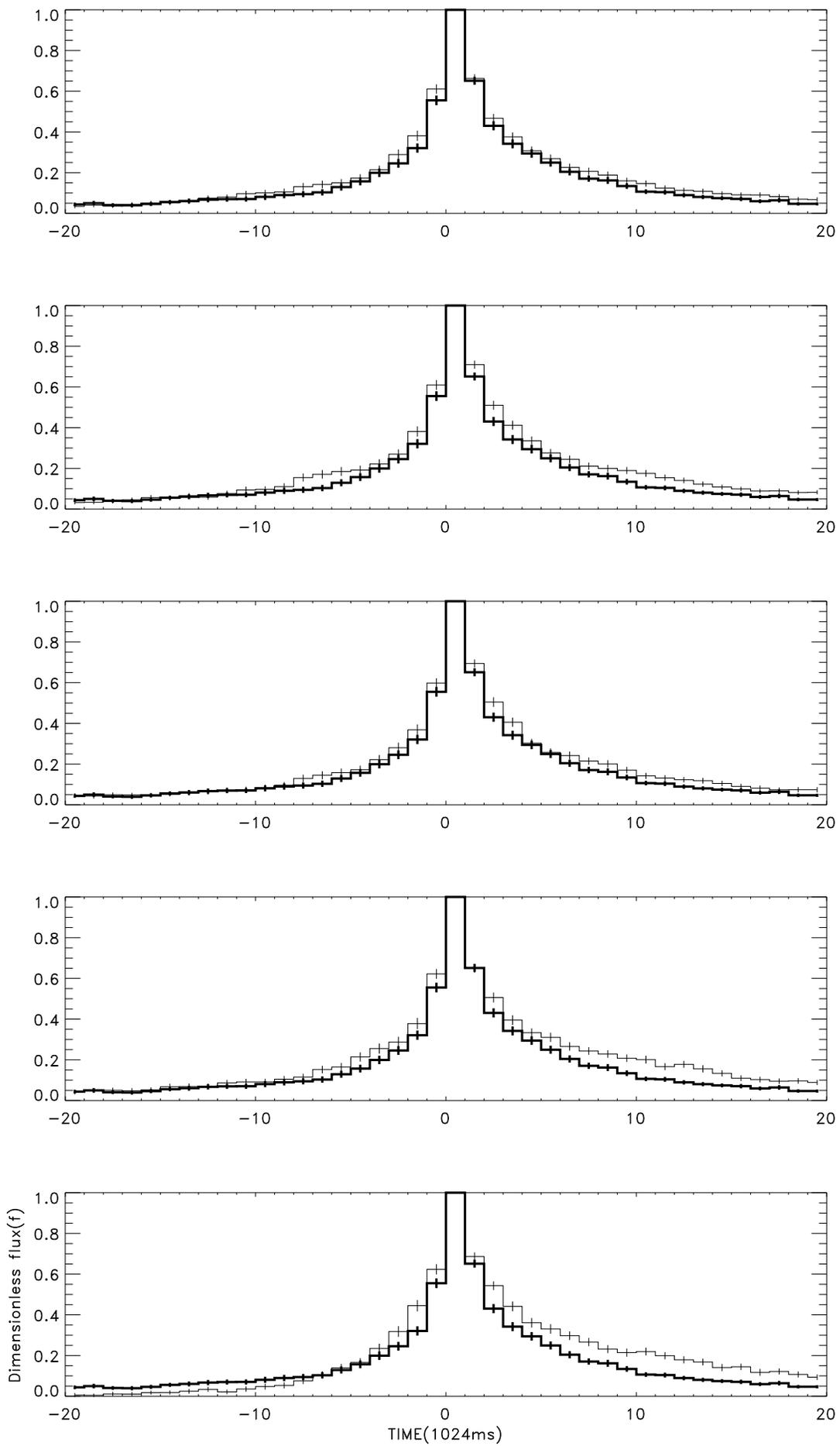
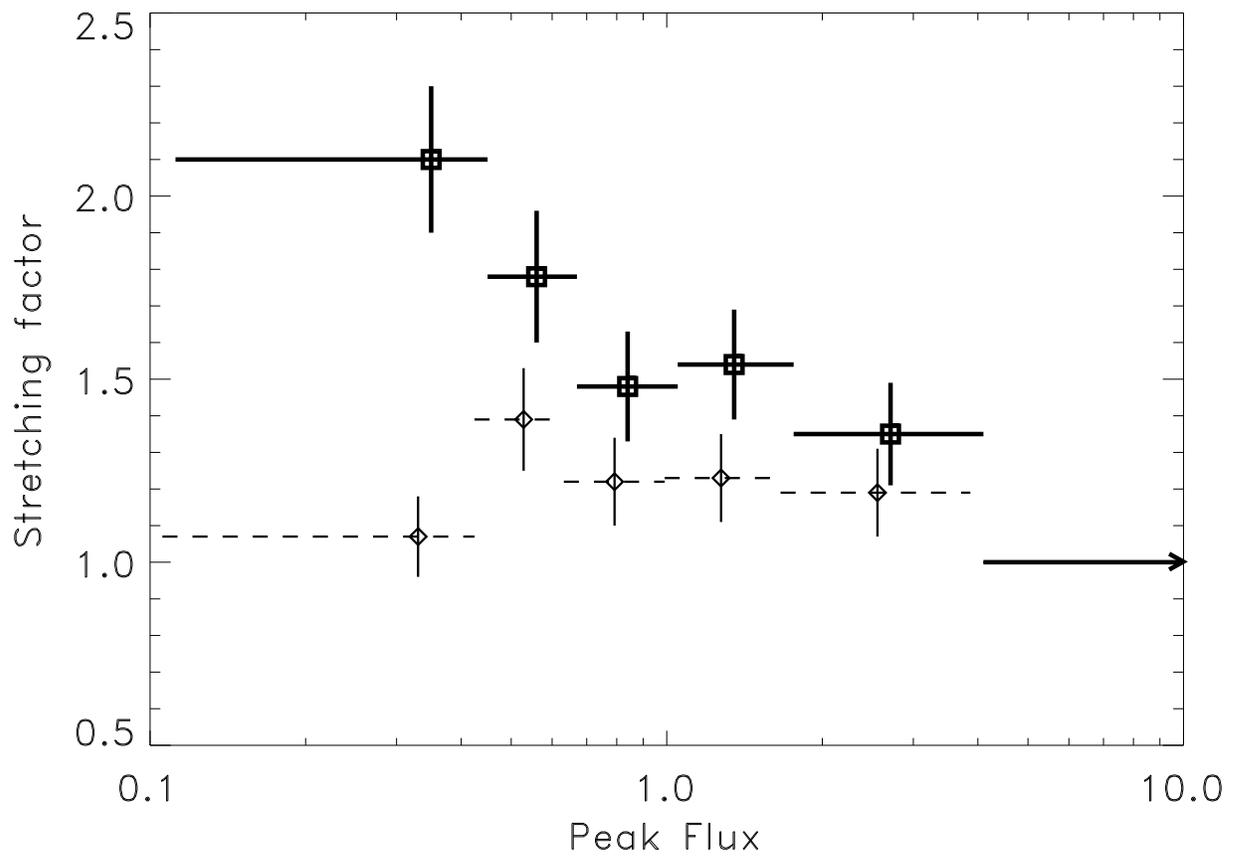
\end{document}